\documentclass[12pt]{article}
\usepackage[cp1251]{inputenc}
\usepackage[english]{babel}
\usepackage[dvips]{graphicx}
\date{}

\begin{document}

\title{The estimation of coherence length for electron-doped superconductor Nd$_{2-x}$Ce$_x$CuO$_{4+\delta}$ }
\author{T.\,B.\,Charikova\/\thanks{charikova@imp.uran.ru}, G.\,I.\,Harus, N.G.~Shelushinina\, O.\,E.\,Sochinskaya\\
{\normalsize\it Institute of Metal Physics RAS, Ekaterinburg, Russia}\\
A.\,A.\,Ivanov\\ {\normalsize\it Moscow Engineering Physics Institute, Moscow, Russia} } \maketitle

\begin{abstract}

Results of low-temperature upper critical field measurements for Nd$_{2-x}$Ce$_x$CuO$_{4+\delta}$ single crystals with various $x$ and nonstoichiometric disorder ($\delta$) are presented. The coherence length of pair correlation $\xi$ and the product $k_F$$\xi$, where $k_F$ is the Fermi wave vector, are estimated. It is shown  that for investigated single crystals parameter $k_F$$\xi$ $\cong$ 100 and thus phenomenologically NdCeCuO - system is in a range of Cooper-pair-based (BCS) superconductivity.


\end{abstract}
\section{Introduction}
In the hole-doped cuprate high-T$_c$ superconductors the size of the pairs, as estimated from the Ginzburg-Landau coherence length $\xi$, is only few times the lattice spacing \cite{randeria} in contrast to ordinary superconductors where the pair size greatly exceeds the lattice spacing or the average distance between carriers. In view of short coherence length of high-T$_c$ superconductors a situation close to compact bosons with Bose-Einstein (BE) condensation at T$_c$ is conceivable. The evolution from BCS superconductivity to BE condensation through the increase of the coupling strength between fermions was studied by Nozieres and Schmitt-Rink \cite{nozieres&schmitt-rink} and it was concluded that the evolution is smooth.

In \cite{pistolesi} convenient phenomenological parameter was selected to establish the crossover from BCS superconductivity to BE condensation of composite bosons, namely, the product $k_F$$\xi$  of Fermi wave vector times the coherence length. Pistolesi et al.  \cite{pistolesi} argued that Cooper-pair-based superconductivity is stable against bosonization down to $k_F$$\xi$ $ =$ 2$\pi$. The stabilization criterion $k_F$$\xi$ $\geq$ 2$\pi$ corresponding to the condition $\xi > \lambda_F$, with $\lambda_F$ $=$ 2$\pi$/$k_F$ being the electron wave length, should be regarded as an analog of the Ioffe-Regel criterion for transport in disordered systems \cite{mott}.

It appears that for hole-doped high-T$_c$ superconductors (series of La-, Y-, Bi- and Tl-systems) $k_F$$\xi$ $\cong$ 10 that are although in a BCS range but near the "instability" line $k_F$$\xi$$ =$ 2$\pi$ on the plot of T$_c$ vs T$_F$($=$ E$_F$/k) of Uemura et al. \cite{uemura}. Our goal was to estimate a parameter $k_F$$\xi$ at electron-doped superconductor Nd$_{2-x}$Ce$_x$CuO$_{4+\delta}$ with various Ce concentration.

\section{Experimental results and discussion}
In order to find $\xi$ the low-temperature measurements of upper critical field B$_{c2}$ on Nd$_{2-x}$Ce$_x$CuO$_{4+\delta}$ single crystal films with various Ce concentration and nonstoichiometric disorder $\delta$ \cite{charikova} in magnetic fields up to 9T ($B \parallel {c}$,  $J \parallel {ab}$) and temperature range 0.4-40\,K with SQUID-magnitometer MPMS XL of Quantum Design and by dc-current method in solenoid up to 12T from “Oxford Instruments” were carried out.

In Fig.1 the dependencies of the resistivity $\rho$ in CuO$_2$ - planes ($J \parallel {ab}$) on perpendicular magnetic field $B \parallel {c}$ are presented for optimally reduced films with $x$ $=$  0.14; 0.15; 0.18; 0.20 and an example of B$_{c2}$ determination ( for $x$ $=$0.15 at T $=$ 0.4 K) is shown. As it should be obtained B$_{c2}$ value is the higest for optimally doped sample with with $x$ $=$ 0.15. 

Fig.2 demonstrates an effect of nonstoichiometric disorder on the upper critical field of optimally doped Nd$_{1.85}$Ce$_{0.15}$CuO$_{4+\delta}$ system. Results of magnetoresistance measurement are presented for three types of Nd$_{2-x}$Ce$_x$CuO$_{4+\delta}$ single crystal films \cite{ivanov}: as-grown samples, optimally reduced samples (optimally annealed in a vacuum at T $=$ 780$^0$\,C for t $=$ 60 min; p $=$ 10$^{-2}$\,mmHg) and non optimally reduced samples (annealed in a vacuum T $=$ 780$^0$\,C for t $=$ 40 min; p $=$ 10$^{-2}$\,mmHg). The film thikness was 1200 - 2000 \AA\ .

Using the relation between the coherence length and the upper critical field   2$\pi$B$_{c2}$$\xi^2 = \Phi_0$ where the elementary flux quantum $\Phi_0 = \pi$$c\hbar$/e, the values of $\xi$ for all samples were estimated. The data for normal state in-plane resistivity and Hall coefficient \cite{charikova} were turned to account for determination of parameter $k_F$$\ell$, mean free path $\ell$ and $k_F$ $=$ (2$\pi$n$_s$)$^{1/2}$, n$_s$ being the surface electron density. All the obtained parameters along with the 
$k_F$$\xi$ values are presented in Table 1 for optimally reduced samples with different Ce concentration  and in Table 2 for samples with $x =$ 0.15 and different nonstoichiometric disorder.
It is known \cite{de Gennes} that for ``dirty'' ($\ell < \xi$) s-wave superconductor 
\begin{equation}
B_{c2}(T=0)= \frac{1}{2\gamma} \cdot \frac{\Phi_0}{\hbar D} \cdot kT_c
\end{equation}
where constant $\gamma \cong $ 1.78, D $=$ $v_F$$\ell$/2 $=$ $\frac {\hbar}{2m}$$k_F \ell$ is the diffusion coefficient, $v_F$ is Fermi velocity. 

Then 
\begin{equation}
\xi = \sqrt{\xi_0 \ell},
\end{equation}
where $\xi_0 \cong \frac {\hbar v_F}{kT_c}$ is the coherence length in pure superconductor. From (1) and (2) we have $B_{c2} \sim (k_F \ell)^{-1}$ and $\xi \sim \sqrt{k_F \ell}$, thus $B_{c2}$ should $increase$ and $\xi$ should $decrease$ with increase of $(k_F \ell)^{-1}$ as a degree of disorder.

As it is seen from Table 2  for Nd$_{1.85}$Ce$_{0.15}$CuO$_{4+\delta}$  the upper critical field quickly $decreases$ and the coherence length  $increases$ with increasing of degree of disorder (parameter $(k_F \ell)^{-1}$) in contradiction with standard results for s-wave superconductor. Such an unusual behavior of $B_{c2}$ and $\xi$ with variation of disorder may be an evidence of d-wave symmetry of superconducting order parameter for Nd$_{1.85}$Ce$_{0.15}$CuO$_{4+\delta}$. It is in accordance with the theoretical considerations of Yin and Maki \cite {Yin} for d-wave superconductors as with our results for a slope of upper critical field in vicinity of $T_c$ in this electron doped system \cite {Charikova1}.

In Fig.3 a log-log plot of $T_c$ versus Fermi temperature $T_F$ $=$ $\varepsilon_F/k$ (so named ``Uemura plot'' \cite{uemura}) for different superconductors is presented and the points for Nd$_{2-x}$Ce$_x$CuO$_{4+\delta}$ system received by us are also shown. The lines with constant $k_F$$\xi$ values ($k_F$$\xi$ $=$ 2$\pi$ and $k_F$$\xi$ $=$ 10$^n$, $n$ $=$ 1$\div$5) are superimposed on the plot according to \cite{pistolesi}. It may be seen that parameter $k_F$$\xi$ $\cong$ 100 for different samples of single crystals Nd$_{2-x}$Ce$_x$CuO$_{4+\delta}$ with various Ce concentration and nonstoichiometric disorder. Thus this electron doped system is even more deep in the region of BCS-coupling than hole-doped cuprate systems. The value of $k_F$$\xi$ is minimal for optimally doped system ( $k_F$$\xi = $ 70$\div$80 and nearly independent on a degree of disorder) and increases for  overdoped ($x$ = 0.18 and 0.20) samples. 

\section{Conclusions}
Thus, from a values of upper critical field we estimate the coherence length in Nd$_{2-x}$Ce$_x$CuO$_{4+\delta}$ system with various $x$ and $\delta$. Then, using the universal (independent of the details of the interaction potential) phenomenological parameter $k_F$$\xi$ \cite{pistolesi}, we illustrate that investigated electron doped cuprate NdCeCuO system  doesn't cross the instability line of BCS superconductivity $k_F$$\xi$ = 2$\pi$ even for optimally doped and optimally reduced samples.

We are grateful to V.N.Neverov for experimental support. This work was done within RAS Programm (project N 01.2.006.13394).

\newpage

\begin{figure}[H!]
	\centering
		\includegraphics[width=1.00\textwidth]{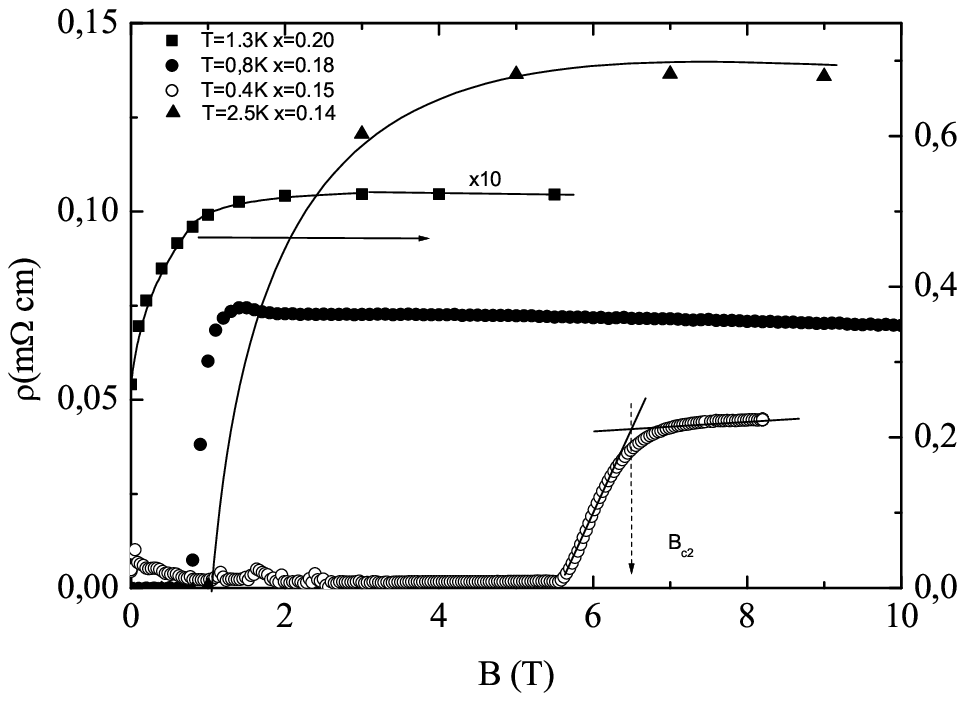}
	\caption{Resistivity at CuO$_2$-planes ($J \parallel {ab}$) vs magnetic field ($B \parallel {c}$) for samples Nd$_{2-x}$Ce$_x$CuO$_{4}$ with different Ce concentration at low temperatures. The  lines are guides to the eye.}
	\label{fig:Fig1}
\end{figure}

\begin{figure}[H!]
	\centering
		\includegraphics[width=1.00\textwidth]{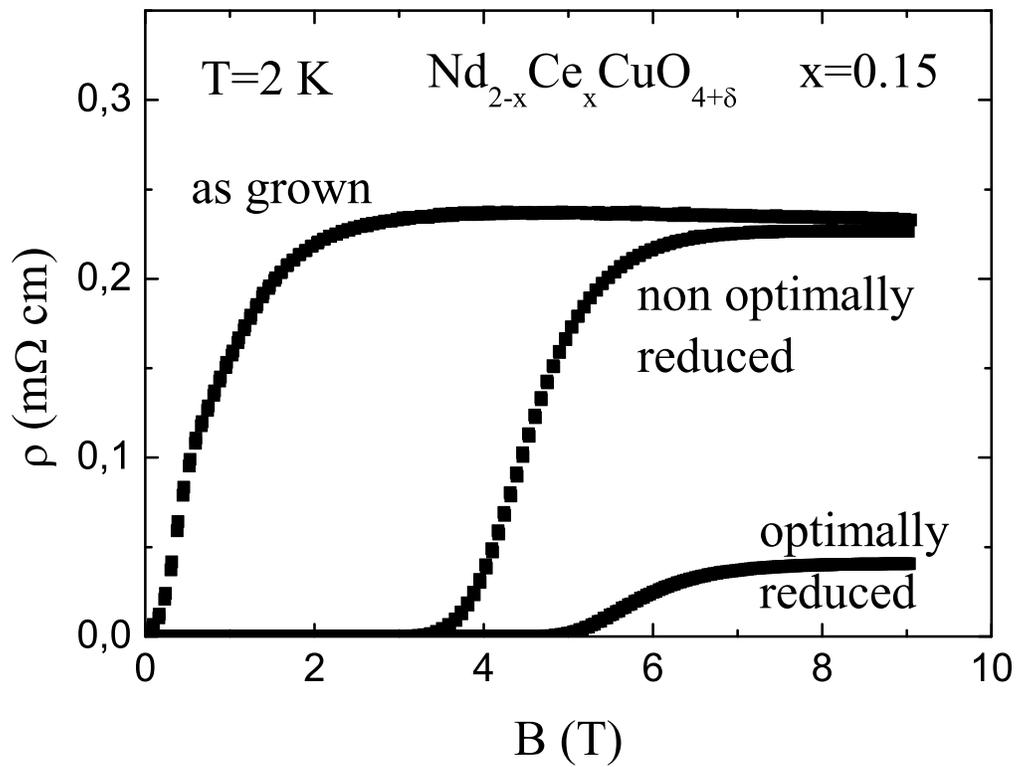}
	\caption{Resistivity at CuO$_2$-planes ($J \parallel {ab}$) vs magnetic field ($B \parallel {c}$) for samples Nd$_{1.85}$Ce$_{0.15}$CuO$_{4}$ with different nonstoichiometric disorder at $T =$ 2 K.}
	\label{fig:Fig2}
\end{figure}

\begin{figure}[H!]
	\centering
		\includegraphics[width=1.00\textwidth]{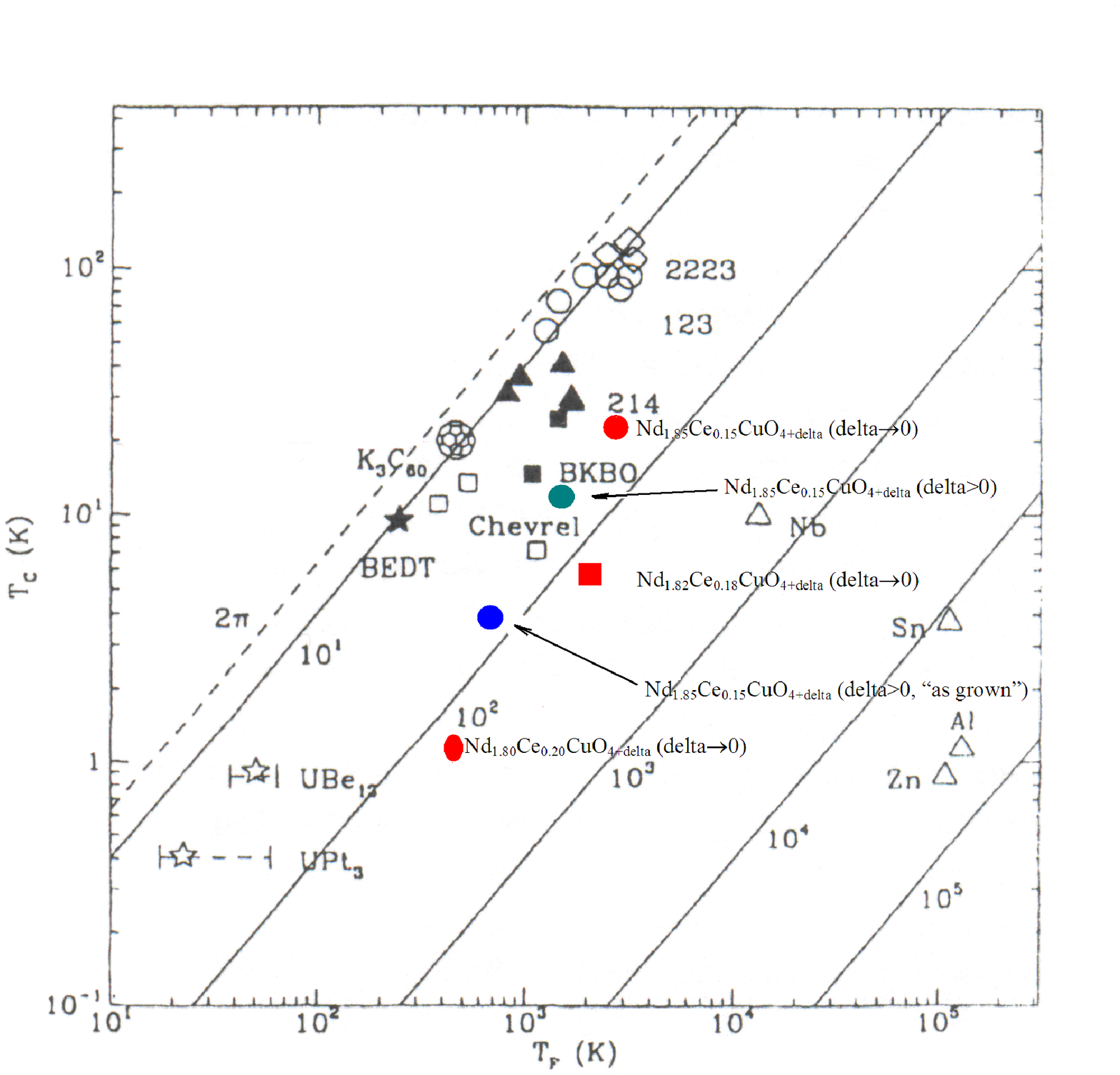}
	\caption{``Uemura plot'' \cite{uemura} with constant $k_F$$\xi$ lines \cite{pistolesi} and with our points for Nd$_{2-x}$Ce$_x$CuO$_{4+\delta}$ system. }
	\label{fig:Fig3}
\end{figure}

\newpage
\section*{Table 1}
\begin{table}[h]
\footnotesize
\caption{The data for Nd$_{2-x}$Ce$_x$CuO$_{4}$ optimally reduced films.}
\label{TBL1}
\begin{center}
\begin{tabular}{|c|c|c|c|c|c|} \hline
Samples& $T_c$,K & $B_{c2}$, T & $k_F$$\ell$ &  $\xi$,\AA\ & $k_F$$\xi$ \\ \hline
x=0.14 & 11 &  2.9 & 2.7 &  106.5 &  - \\ \hline  
x=0.15 & 21  & 6.1 & 51.6 & 73.5 &  74.2 \\ \hline
x=0.18 & 6 & 0.76 & 44.4  & 207.7 & 118.4 \\ \hline
x=0.20 & $<$1.3 & 0.4 & 14.6 & 273.3 & 166.7 \\ \hline
\end{tabular}
\end{center}
\end{table}

\section*{Table 2}
\begin{table}[h]
\footnotesize
\caption{The data for Nd$_{1.85}$Ce$_{0.15}$CuO$_{4+\delta}$ films with different nonstoichiometric disorder.}
\label{TBL2}
\begin{center}
\begin{tabular}{|c|c|c|c|c|c|} \hline
Samples& $B_{c2}$, T & $k_F$$\ell$ & $\ell$,\AA\ & $\xi$,\AA\ & $k_F$$\xi$ \\ \hline
Optimally reduced & 6.1  & 51.6 & 51.3 &73.5 & 74.2 \\ \hline
Non optimally reduced& 4.8 & 9.1 & 12.5  & 82.3 & 68.3 \\ \hline
As grown & 1.3 & 8.6 & 13.4 & 158.7 & 80.9 \\ \hline
\end{tabular}
\end{center}
\end{table}
\end{document}